\documentclass[conference,letterpaper]{IEEEtran}

\usepackage[utf8]{inputenc} 
\usepackage[T1]{fontenc}
\usepackage{url}
\usepackage{ifthen}
\usepackage{cite}
\usepackage[cmex10]{amsmath} 

\usepackage{physics}
\usepackage{amsfonts, amssymb}
\usepackage{graphicx}
\ifCLASSOPTIONcompsoc
    \usepackage[caption=false,font=normalsize,labelfont=sf,textfont=sf{subfig}
\else
    \usepackage[caption=false,font=footnotesize]{subfig}
\fi
\usepackage[inline, shortlabels]{enumitem}

\newtheorem{theorem}{Theorem}[section]
\newtheorem{corollary}[theorem]{Corollary}
\newtheorem{proposition}[theorem]{Proposition}
\newtheorem{rem}{Remark}

\usepackage{xcolor}

\begin{document}
\title{Quantum Entropy Prover} 


\author{%
    \IEEEauthorblockN{Shao-Lun Huang}
    \IEEEauthorblockA{Institute for Quantum Information\\
                    RWTH Aachen University\\
                    Germany\\ 
                    Email: shao-lun.huang@outlook.com
                    }
    \and 
    \IEEEauthorblockN{Tobias Rippchen}
    \IEEEauthorblockA{Institute for Quantum Information\\
                    RWTH Aachen University\\
                    Germany\\
                    Email: tobias.rippchen@rwth-aachen.de
                    }
    \and
    \IEEEauthorblockN{Mario Berta}
    \IEEEauthorblockA{Institute for Quantum Information\\
                    RWTH Aachen University\\
                    Germany\\ 
                    Email: berta@physik.rwth-aachen.de
                    }   
}

\maketitle


\begin{abstract}   
    Information inequalities govern the ultimate limitations in information theory and as such play an pivotal role in characterizing what values the entropy of multipartite states can take. Proving an information inequality, however, quickly becomes arduous when the number of involved parties increases. For classical systems, [Yeung, IEEE Trans.\ Inf.\ Theory (1997)] proposed a framework to prove Shannon-type inequalities via linear programming. Here, we derive an analogous framework for quantum systems, based on the strong sub-additivity and weak monotonicity inequalities for the von-Neumann entropy. Importantly, this also allows us to handle constrained inequalities, which\,---\,in the classical case\,---\,served as a crucial tool in proving the existence of non-standard, so-called non-Shannon-type inequalities [Zhang \& Yeung, IEEE Trans.\ Inf.\ Theory (1998)]. Our main contribution is the Python package qITIP, for which we present the theory and demonstrate its capabilities with several illustrative examples.
\end{abstract}


\section{Introduction} \label{sec:intro}




Information inequalities are inequalities that can be expressed as linear combinations of the entropy of multipartite states. Since such inequalities govern what is ultimately feasible in information theory, they are also regarded as the laws of information theory \cite{Yeung} and play a vital role in characterizing the entropy and other information measures (e.g.\ \cite{Zhang_1998, Pippenger_2003, Matus_2007, Linden_2005, Majenz_2018}). However, a challenge is that the proof of an information inequality is usually not mathematically obvious. In particular, in quantum information theory, the only inequalities currently known to be satisfied by all quantum states can be derived from the strong sub-additivity (SSA) and weak monotonicity (WM) of the quantum entropy \cite{Majenz_2018}. In the literature, these are called \emph{von-Neumann-type inequalities}. A quantum information inequality that is not von-Neumann-type, but nevertheless satisfied by all quantum states would accordingly be called a \emph{non-von-Neumann-type inequality}. Currently, the existence of non-von-Neumann-type inequalities remains an open question \cite{Linden_2005, Cadney_2012, Majenz_2018}. 


In classical information theory, a linear programming approach is used to prove the classical counterparts to von-Neumann-type inequalities (cf.\ \cite{Yeung_1997, Ho_2020, Yeung}). Namely, the so-called \emph{Shannon-type inequalities} can be derived by SSA and strong monotonicity (SM) of the classical entropy function. SM is equivalent to the non-negativity of the classical conditional entropy and it is this fact that is one of the biggest differences between classical and quantum information theory. The quantum conditional entropy can be negative; hence, a different set of basic inequalities is required here. In \cite{Pippenger_2003}, Pippenger found the minimal set of inequalities that generate all instances of SSA and WM.

Inspired by the classical prover, our work combines Pippenger's minimal set with the linear programming technique by Yeung. Our main contribution is a Python package based on our framework, called qITIP \cite{qitip_2024}, that automates the proof of von-Neumann-type inequalities. Moreover, the existence of \emph{non-Shannon-type inequalities} was answered in the affirmative in \cite{Zhang_1998} after a constrained one was found in \cite{Zhang_1997}. Our framework then also includes the possibility to handle additional constraints to facilitate similar research into non-von-Neumann-type inequalities. 




\section{Quantum Entropic Cone}

A quantum state \(\hat{\rho}\) is a unit-trace positive semi-definite operator on a finite-dimensional Hilbert space \(\mathcal{H}\). The entropy of $\hat{\rho}$, the so-called quantum entropy or von-Neumann entropy, is defined as
\begin{equation}
\label{eq:def_quantum_entropy}
S(\hat{\rho}) \equiv -\Tr[\hat{\rho}\log{\hat{\rho}}]
\end{equation}
with the trace operation \(\Tr[\cdot]\). 

For an \(n\)-party quantum system, the Hilbert space factors into the tensor product \( \mathcal{H} = \bigotimes_{i=1}^n \mathcal{H}_i \) of the individual Hilbert spaces \( \mathcal{H}_i\). In the following, we will often denote the parties by capital letters \(A,B,C,\text{etc}\).  The state of an \(n\)-party system will be represented as $\hat{\rho}_{N_n}$, where $N_{n}$ abbreviates the set $\{1,2,\dots,n\}$. If we focus on the subsystems $I \subseteq N_{n}$, the marginal quantum state $\hat{\rho}_{I}$ is given by
\begin{equation}
\hat{\rho}_{I} \equiv \Tr_{N_{n}\setminus I}[\hat{\rho}_{N_{n}}],
\end{equation}
where $\Tr_{N_{n}\setminus I}[\cdot]$ means partially tracing out the parts not included in $I$. Therefore, given a $n$-party quantum state $\hat{\rho}_{N_n}$, there are $k = 2^{n} - 1$ marginal states associated with it excluding the trivial null state $\hat{\rho}_{\emptyset}$. The $k$-dimensional vector whose entries are given by these $k$ marginal entropies is called the \emph{entropic vector}. The associated vector space $\mathbb{R}^{k}$ is called the \emph{entropic space}.  

Consider, for example, a bipartite quantum system whose quantum state is $\hat{\rho}_{AB}$. The entropic vector of $\hat{\rho}_{AB}$ is given by
\begin{equation}
\mathbf{s}(\hat{\rho}_{AB}) = \langle{S(\hat{\rho}_{A}), S(\hat{\rho}_{B}), S(\hat{\rho}_{AB})}\rangle.
\end{equation}
We use $\mathbf{s}(\hat{\rho}_{AB})$ to indicate the entropic vector of the state $\hat{\rho}_{AB}$. For ease of notation, we will drop the symbol $\hat{\rho}$ for the marginal entropies, leaving only the indices of the subsystems, e.g., \(S(A) \equiv  S(\hat{\rho}_{A})\).

In the entropic space, let $\Gamma_{n}^{*}$ denote the set of all entropic vectors associated with a valid \(n\)-party quantum state. The closure of $\Gamma_{n}^{*}$, denoted by $\overline{\Gamma_{n}^{*}}$, was proven by Pippenger to form a convex cone called the \emph{entropic cone} in \cite{Pippenger_2003}. He then initiated a program to characterize the structure of this entropic cone using information inequalities.


\subsection{Von-Neumann Cone}

First, let us formally introduce the two most important information inequalities in quantum information theory\,--\,the strong sub-additivity (SSA),
\begin{equation}
S(I) + S(J) \geq S(I \cup J) + S(I \cap J),
\end{equation}
and the weak monotonicity (WM),
\begin{equation}
S(I) + S(J) \geq S(I \setminus J) + S(J \setminus I),
\end{equation}
where $I, J\subseteq N_{n}\setminus\emptyset$. The most familiar instances of SSA and WM are probably those of a tripartite system (see \cite{Lieb_1973} for the original proof) given by
\begin{equation}
\label{eq:cond_mutual_expression}
    S(A, B) + S(B, C) \geq S(A, B, C) + S(B)
\end{equation}
and 
\begin{equation}
    S(A, B) + S(B, C) \geq S(A) + S(C).
\end{equation}
Note that equation \eqref{eq:cond_mutual_expression} can be re-expressed as the positivity of the quantum conditional mutual information defined as
\begin{equation}
    I(A;C\vert B) \equiv S(A, B) + S(B, C) - S(A, B, C) + S(B).
\end{equation}

Note that although WM can be derived from SSA, the proof requires an additional party to which we have no access in our setting. We refer to all instances of SSA and WM as \emph{basic inequalities}. As we can see, the basic inequalities are given in terms of marginal entropies; therefore, we can write all of them compactly in matrix form as $\mathbf{\Tilde{G}}\mathbf{s} \geq \mathbf{0}$, where $\mathbf{0}$ is the null vector and $\mathbf{\Tilde{G}}$ is a matrix derived from the coefficients in front of the marginal entropies. 

\begin{figure}
    \centering
    \subfloat[$\overline{\Gamma_{n}^{*}} \subsetneq \Gamma_{n}$]
    {
        \centering
        \includegraphics[width=2.5in]{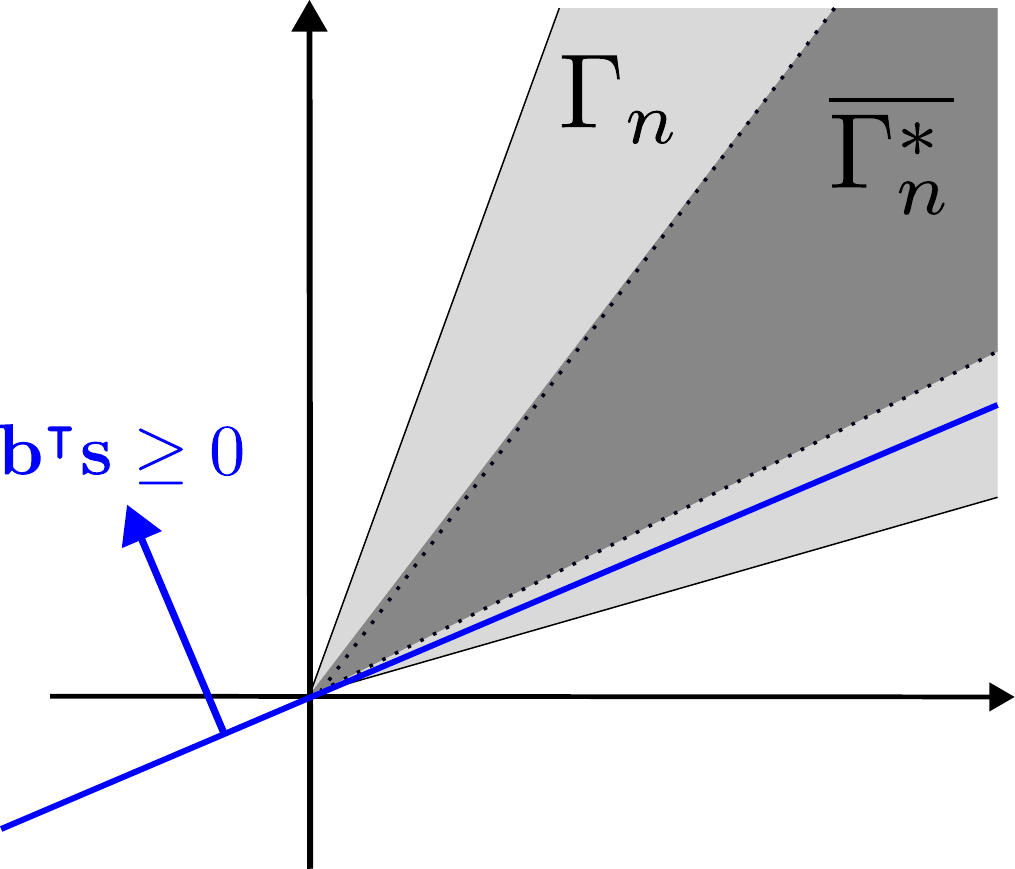}
        \label{fig:non-von-neumann}
    } \\
    \subfloat[$\overline{\Gamma_{n}^{*}} = \Gamma_{n}$]
    {
        \centering
        \includegraphics[width=2.5in]{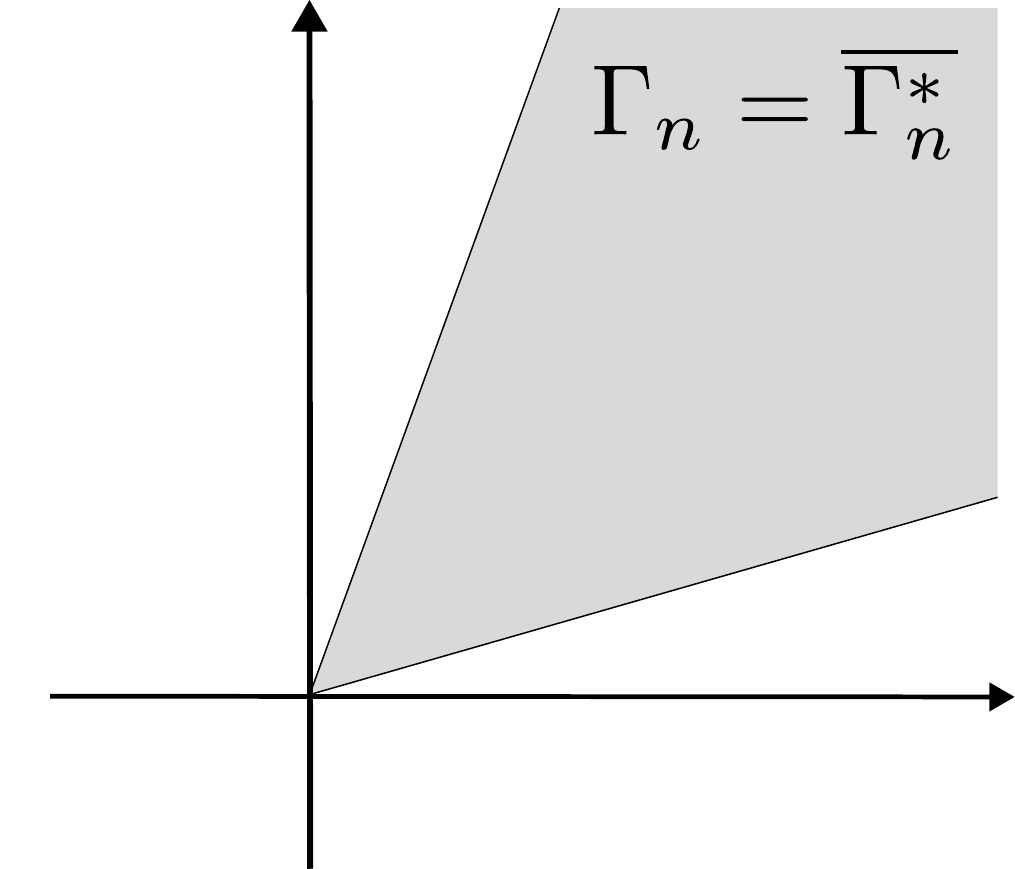}
        \label{fig:only-von-neumann}
    }
    \caption
    {
    Implications of (non)-existence of non-von-Neumann-type inequalities. \protect\subref{fig:non-von-neumann} There exists an information inequality $\mathbf{b^{\top}\mathbf{s}} \geq 0$ such that every element in $\overline{\Gamma_{n}^{*}}$ satisfies the inequality. However, we see there are elements within $\Gamma_{n}$ that do not. \protect\subref{fig:only-von-neumann} Since $\Gamma_{n} = \overline{\Gamma_{n}^{*}}$, if every element in $\overline{\Gamma_{n}^{*}}$ satisfies $\mathbf{b^{\top}\mathbf{s}} \geq 0$, so does every element in $\Gamma_{n}$.
    }
    \label{fig:cone}
\end{figure}

In entropic space, we then define another set $\Gamma_{n}\subset\mathbb{R}^{k}$ as
\begin{equation}
    \Gamma_{n} = \left\{\mathbf{s}\in\mathbb{R}^{k} \colon \mathbf{\Tilde{G}}\mathbf{s} \geq \mathbf{0} \right\},
\end{equation}
which we call the \emph{von-Neumann cone}. Since all quantum states satisfy SSA and WM, we have
\begin{equation}
\label{eq:set-relation}
    \Gamma_{n}^{*} \subset \overline{\Gamma_{n}^{*}} \subset \Gamma_{n}.
\end{equation}
An open question is whether $\Gamma_{n} = \overline{\Gamma_{n}^{*}}$ holds for \(n\geq4\) \cite{Pippenger_2003}. If this were not the case, it implies the existence of non-von-Neumann-type inequalities. A visualization of the argument is presented in Fig.\ \ref{fig:cone}.


\subsection{Pippenger's Minimal Set of Inequalities}

Note that some basic inequalities can be derived from other basic inequalities. For example, the inequality
\begin{equation}
    I(A; B, C) \equiv S(A) + S(B,C) - S(A,B,C) \geq 0
\end{equation}
can be derived from $I(A; B) \geq 0$ and $I(A; C\vert B) \geq 0$ combined with the chain rule
\begin{equation}
   I(A; B, C) = I(A; B) + I(A; C\vert B). 
\end{equation}
Pippenger \cite{Pippenger_2003} showed that the basic inequalities can be reduced to a set of linearly independent inequalities that is complete. We will call this the set of \emph{elemental inequalities}. In the following, let $I, J\subseteq N_{n}$ and \(1 \leq i,j,k \leq n\). The elemental inequalities consist of the following two sets:
\begin{enumerate}[label=(\roman*)]
    \item $S(I) + S(J) - S(I\cup J) - S(I\cap J) \geq 0$ where $I\setminus J=\{i\}$, $J\setminus I = \{j\}$ and $j > i$
    \item $S(I) + S(J) - S(I\setminus J) - S(J\setminus I) \geq 0$ where $I\cap J = \{k\}$, $I\cup J = N_{n}$ and $k + 1 \in I$.
\end{enumerate}
Analogous to the basic inequalities, we can express the elemental inequalities compactly as $\mathbf{Gs} \geq \mathbf{0}$ wit a suitable coefficient matrix \(\mathbf{G}\). Therefore, the von-Neumann cone can equivalently be expressed as
\begin{equation}
    \Gamma_{n} = \{\mathbf{s}\in\mathbb{R}^{k} \colon \mathbf{Gs} \geq \mathbf{0} \}.
\end{equation}


\section{Information-Theoretic Inequality Prover}


We will now present a mathematical framework based on Pippenger's characterization of the von-Neumann cone that can automate the proof of linear information inequalities. We implemented this framework in a Python package called qITIP. Its code and documentation is available on GitHub \cite{qitip_2024}.


\subsection{Framework of qITIP}

Consider a particular linear information inequality given by
\begin{equation}
\label{eq:def-von-neumann}
    \mathbf{b}^{\top}\mathbf{s}\geq 0
\end{equation}
with associated coefficient vector \(\mathbf{b}\).
If 
\begin{equation}
\label{eq:von-neumann-type}
    \Gamma_{n}\subset\{\mathbf{s}\in\mathbb{R}^{k}\colon \mathbf{b}^{\top}\mathbf{s}\geq 0 \}
\end{equation}
holds, we know $\overline{\Gamma_{n}^{*}} \subset \{\mathbf{s}\in\mathbb{R}^{k}\colon \mathbf{b}^{\top}\mathbf{s}\geq 0 \}$ by Equation \eqref{eq:set-relation}. Von-Neumann-type inequalities are exactly those inequalities that satisfy Equation \eqref{eq:von-neumann-type}. Similar to the classical case, we have the following theorem for unconstrained von-Neumann-type inequalities (cf.\ \cite{Yeung_1997}).
\begin{theorem}
\label{thm:primal_unconstrained}
    With definitions as above, the inequality $\mathbf{b}^{\top}\mathbf{s}\geq 0$ is von-Neumann-type if and only if
    \begin{equation}
    \label{eq:primal_unconstrained}
        \min_{\mathbf{s}\in\mathbb{R}^{k}}\mathbf{b}^{\top}\mathbf{s} \, \, \text{ subject to }\mathbf{Gs \geq 0}
    \end{equation}
    is zero.
\end{theorem}
\begin{IEEEproof}
    Assume $\mathbf{b}^{\top}\mathbf{s}\geq 0$ is von-Neumann-type thus $\mathbf{b}^{\top}\mathbf{s}\geq 0$ holds for all \(\mathbf{s} \in\overline{\Gamma_{n}^{*}}\). Since a product state of pure states corresponds to the origin in the entropic space, we then have
    \begin{equation}
        \min_{\mathbf{s}\in\mathbb{R}^{k}} \left\{ \mathbf{b}^{\top}\mathbf{s} \colon \mathbf{Gs \geq 0} \right\} = 0 .
    \end{equation}
    Conversely, assume that
    \begin{equation}
        \min_{\mathbf{s}\in\mathbb{R}^{k}} \left\{ \mathbf{b^{\top}s} \colon \mathbf{Gs \geq 0} \right\} = 0 .
    \end{equation}
    By contradiction, if $\Gamma_{n}\not\subset \{\mathbf{s}\in\mathbb{R}^{k}\colon \mathbf{b}^{\top}\mathbf{s}\geq 0 \}$ there exists some $\mathbf{s}^{\prime}\in\Gamma_{n}$ such that $\mathbf{b}^{\top}\mathbf{s}^{\prime} < 0$ which violates the assumption.
\end{IEEEproof}

If we view Theorem \ref{thm:primal_unconstrained} as the "geometric" perspective on the problem, we can also tackle it from an "algebraic" perspective using duality of linear programming. 

\begin{corollary}
    With definitions as above, the inequality $\mathbf{b}^{\top}\mathbf{s}\geq 0$ is von-Neumann-type if and only if there exists a solution $\mathbf{y}^{*}$ to the problem
    \begin{equation}
    \label{eq:dual-unconstrained}
        \mathbf{y}^{\top}\mathbf{G} = \mathbf{b}^{\top} \textrm{ and }\mathbf{y \geq 0}.
    \end{equation}
\end{corollary}

We omit the proof as transforming a linear program into its dual form follows from textbook techniques (cf.\ e.g.\ \cite[Chapter 5]{Boyd}).




\begin{rem}
    Note that if an inequality $\mathbf{b^{\top}s} \geq 0$ is not von-Neumann-type, the answer to equation \eqref{eq:primal_unconstrained} is $-\infty$. This can be seen as follows. Consider an entropic vector $\mathbf{s'}\in \Gamma_{n}$ such that $\mathbf{b^{\top}s'} < 0$ and $\mathbf{Gs' \geq 0}$ and let $\lambda\in \mathbb{R}$ with $\lambda > 1$. We then know that 
    \begin{equation}
        \mathbf{G}(\lambda\mathbf{s'}) \geq \mathbf{0} \; \text{and} \; \mathbf{b}^{\top}(\lambda\mathbf{s'}) < \mathbf{b}^{\top}\mathbf{s'} < 0 .
    \end{equation}
\end{rem}


\subsection{Adding Constraints}

In information theory, we often deal with information inequalities that only hold under additional constraints such as the Markov chain condition $I(A;C\vert B) = 0$. In the classical case, for example, a constrained non-Shannon-type inequality was found before the existence of an unconstrained one was established \cite{Zhang_1997}. Moreover, only constrained non-von-Neumann-type inequalities are known so far \cite{Linden_2005, Cadney_2012, Majenz_2018}. In the following, we discuss how constraints can be added into our framework.

Suppose we have $q$ constraints that are expressed as
\begin{equation}
    \mathbf{Qs = 0},
\end{equation}
where each row of $\mathbf{Q}$ represents a constraint written in terms of marginal entropies. Therefore, whether an information inequality is always true under the constraints $\mathbf{Qs = 0}$ can be translated into asking whether the following relation is true:
\begin{equation}
\label{eq:geometric_constrained}
    \overline{\Gamma_{n}^{*}} \cap \{\mathbf{s}\colon \mathbf{Qs = 0}\}  \subset \{\mathbf{s}\in \mathbb{R}^{k}\colon \mathbf{b}^{\top}\mathbf{s} \geq 0\}.
\end{equation}
Using a similar argument as above, we know if
\begin{equation}
\label{eq:constrained_von_neumann}
    \Gamma_{n}\cap \{\mathbf{s}\colon \mathbf{Qs = 0}\}  \subset \{\mathbf{s}\in \mathbb{R}^{k}\colon \mathbf{b}^{\top}\mathbf{s} \geq 0\}
\end{equation}
is true, Equation \eqref{eq:geometric_constrained} is also true. Such inequalities are termed \emph{constrained von-Neumann-type inequalities}.

\begin{proposition}
\label{thm:primal_constrained}
With definitions as above, $\mathbf{b^{\top}s} \geq 0$ is a von-Neumann-type inequality under the constraints $\mathbf{Qs = 0}$ if and only if 
\begin{equation}
\label{eq:primal_constrained}
\min_{\mathbf{s}\in \mathbb{R}^{k}}{\mathbf{b^{\top}s}} \, \, \text{ subject to }\mathbf{Gs\geq 0} \text{ and }\mathbf{Qs = 0}
\end{equation}
is equal to $0$.
\end{proposition}

\begin{IEEEproof}[Proof Idea]
    Since under the constraints $\mathbf{Qs = 0}$, we only need to consider entropic vectors in the kernel of $\mathbf{Q}$, the constrained problem is equivalent to an unconstrained one on the subspace induced by the kernel. Working in this subspace, the proof is essentially the same as for Theorem \ref{thm:primal_unconstrained}. 
\end{IEEEproof}

\begin{rem}
    Analogous to the unconstrained case, if an inequality $\mathbf{b^{\top}s} \geq 0$ is not von-Neumann-type under the constraints $\mathbf{Qs = 0}$, the answer to Equation \eqref{eq:primal_constrained} is $-\infty$. 
\end{rem}

As before, we can also analyse the problem from an algebraic perspective by considering the dual of the above problem.

\begin{proposition}
\label{thm:dual_constrained}
With definitions as above, $\mathbf{b}^{\top}\mathbf{s}$ is von-Neumann-type under the constraints $\mathbf{Q}\mathbf{s} = 0$ if and only if there exists an optimal solution $[\mathbf{y}^{\top}, \mathbf{\mu}^{\top}]$ such that 
\begin{equation}
\label{eq:dual_constrained}
\mathbf{b}^{\top} = \mathbf{y}^{\top}\mathbf{G} - \mathbf{\mu}^{\top}\mathbf{Q} \textrm{ and } \mathbf{y}\geq 0 .
\end{equation}
\end{proposition}





\subsection{Generating Counterexamples}

Inequalities that are not von-Neumann-type could still be valid information inequalities due to the possible existence of non-von-Neumann-type inequalities. A challenge is that there is not a one-to-one correspondence between entropic vectors and quantum states. Therefore, it is hard to construct a counterexample directly from some vector $\mathbf{s}$ in entropic space. However, as we shall see, the dual problem can help disprove inequalities that are not von-Neumann-type by providing a set of sufficient conditions that a potential counterexample has to satisfy. This gives a heuristic way to disprove an information inequality.

As we have shown above, if an information inequality is not von-Neumann-type, the solution to Equation \eqref{eq:primal_constrained} is infinite and no optimal solution to Equation \eqref{eq:dual_constrained} can be found. Therefore, to find a counterexample, we bound the search region in entropic space further by adding more constraints. By requiring $S(\{x\}) \leq 1$ for all $x\in N_{n}$, we have set upper bounds for all $I\subseteq N_{n}$. This can be shown using SSA, which implies
\begin{equation}
S(I) + S(J) \geq S(I\cup J)
\end{equation}
with $I, J\subseteq N_{n}$ and $I\cap J = \emptyset$. More specifically, $S(I)\leq \left|I\right|$ where $\left|I\right|$ is the number of elements in $I$. These constraints can be expressed as $\mathbf{Ws\leq 1}$, where $\mathbf{1}$ is the vector with all entries being 1 and $\mathbf{W}$ is a $n\times k$ matrix with $\mathbf{W} = [\mathbf{I}, \mathbf{\tilde{O}}]$ in which $\mathbf{I}$ is an $n\times n$ identity matrix and $\mathbf{\tilde{O}}$ is an $n\times (k-n)$ null matrix. By imposing these extra constraints, the minimum value of the primal problem in Proposition \ref{thm:primal_constrained} is finite. The primal problem is transformed into
\begin{equation}
\label{eq:primal_disprove}
\min_{\mathbf{s}\in\mathbb{R}^{k}} \mathbf{b^{\top}s} \, \, \text{ subject to }\mathbf{Gs\geq 0}, \mathbf{Qs = 0} \text{ and }\mathbf{Ws \leq \mathbf{1}}.
\end{equation}
The corresponding dual problem is
\begin{equation}
\label{eq:dual_disprove}
\max_{\mathbf{\lambda}\in \mathbb{R}^{n}} -\mathbf{\lambda^{\top}1} \text{subject to}
\begin{cases}
\mathbf{b^{\top}} = \mathbf{y^\top G} - \mathbf{\mu^{\top} Q} - \mathbf{\lambda^{\top} W}\\
\mathbf{y \geq 0}\\
\mathbf{\lambda \geq 0}
\end{cases}.
\end{equation}
As mentioned above, finding a quantum state given an entropic vector is not straightforward. All we can do is try to identify the properties that such a quantum state must satisfy based on the dual problem.

\begin{proposition}
\label{prop:disprove}
With definitions as above, let $[\mathbf{y}^{*\top}, \mathbf{\mu}^{*\top}, \mathbf{\lambda}^{*\top}]$ be an optimal solution to Equation \eqref{eq:dual_disprove}. If there exists a quantum state whose entropic vector $\mathbf{s'}$ satisfies
\begin{equation}
\label{eq:conditions}
\begin{cases}
\mathbf{y}^{*\top}\mathbf{G}\mathbf{s'} = 0\\ 
\mathbf{Q}\mathbf{s'} = 0\\ 
0 < \mathbf{W}\mathbf{s}\leq \mathbf{1}\\
\end{cases},
\end{equation}
then $\mathbf{b}^{\top}\mathbf{s'} < 0$. This shows that the state is a counterexample to the constrained information inequality.
\end{proposition}

\begin{IEEEproof}
The optimal solution, $[\mathbf{y}^{*\top}, \mathbf{\mu}^{*\top}, \mathbf{\lambda}^{*\top}]$, satisfies
\begin{equation*}
\mathbf{b}^{\top} = \mathbf{y}^{*\top}\mathbf{G} - \mathbf{\mu}^{*\top}\mathbf{Q} - \mathbf{\lambda}^{*\top}\mathbf{W}.
\end{equation*}
If we have $\mathbf{b}^{\top}$ acting on $\mathbf{s'}$, then
\begin{equation}
\label{eq:prop_calculation}
\begin{aligned}
\mathbf{b}^{\top}\mathbf{s'} 
&=  \mathbf{y}^{*\top}\mathbf{G}\mathbf{s'} - \mathbf{\mu}^{*\top}\mathbf{Q}\mathbf{s'} - \mathbf{\lambda}^{*\top}\mathbf{W}\mathbf{s'}\\
& = - \mathbf{\lambda}^{*\top}\mathbf{W}\mathbf{s'}.
\end{aligned}
\end{equation}
Combining $\mathbf{W}\mathbf{s'} \leq \mathbf{1}$, $\mathbf{W}\mathbf{s'} \neq 0$ and $\mathbf{\lambda}^{*}\geq 0$, we know $\mathbf{b}^{\top}\mathbf{s'}$ is negative as long as $\mathbf{\lambda}^{*}\neq 0$ which is always the case.
\end{IEEEproof}

At this point, it cannot be overemphasized that a counterexample may not be found despite the hints provided by the solver.


\section{Shortest Proof and Shortest Counterexamples}
\label{sec:shortest_proof}

In general, the solution to the dual problem is not unique. For example, the quantum counterpart of a classical example provided by Ho et.\ al.\ \cite{Ho_2020} is
\begin{equation*}
\begin{aligned}
&S(A,B, C) - S(A\vert B, C) - S(B\vert A, C) - S(3\vert A,B) \\
=& S(A, B) + S(B, C) + S(A, C) - 2\cdot S(A,B,C)\\
=&I(A;B\vert C) + I(A; C) + I(B;C\vert A),
\end{aligned}
\end{equation*}
which can also be rewritten as 
\begin{equation*}
\begin{split}
&S(A,B, C) - S(A\vert B, C) - S(B\vert A, C) - S(C\vert A,B) \\
=&0.8[I(A;B\vert C) + I(A; C) + I(B;C\vert A)] + \\
&0.1[I(B;C\vert A) + I(A; B) + I(A;C\vert B)] + \\
&0.1[I(A; C\vert B) + I(B; C) + I(A;B\vert C)].
\end{split}
\end{equation*} 
This is clearly not as clean and straightforward as the previous formulation. In this section, we show how the prover can generate the shortest proofs.


\subsection{Generating Shortest Proof}

If a given quantum inequality proved to be von-Neumann-type, its shortest proof is considered one that requires the least number of elemental inequalities and the least number of constraints. Mathematically, this requirement can be written as 
\begin{equation}
\min_{\substack{\mathbf{y}\in \mathbb{R}^{m},\\ \mathbf{\mu}\in \mathbb{R}^{q}}} \| [\mathbf{y^{\top}}, \mathbf{\mu^{\top}}]\|_{0} \text{ subject to }
\begin{cases}
    \mathbf{y^{\top}G} = \mathbf{b}^{\top} + \mathbf{\mu^{\top}Q}\\
    \mathbf{y}\geq 0
\end{cases},
\end{equation}
where $\|\mathbf{v} \|_{0}$ is the \emph{$\mathnormal{l}_{0}$-norm}. This is generally a computationally hard problem, so instead we solve a relaxation involving the \emph{$\mathnormal{l}_{1}$-norm} instead:
\begin{equation}
\label{eq:crude_min_proof}
\min_{\substack{\mathbf{y}\in \mathbb{R}^{m},\\ \mathbf{t}\in \mathbb{R}^{q}}} \mathbf{1^{\top}y} + \| \mathbf{\mu} \|_{1} \text{ subject to }
\begin{cases}
    \mathbf{y^{\top}G} = \mathbf{b}^{\top} + \mathbf{\mu^{\top}Q}\\
    \mathbf{y}\geq 0
\end{cases}.
\end{equation}
This problem is obviously not equivalent to the $\mathnormal{l}_{0}$ norm problem. However, it gives a good approximation; intuitively, as the $\mathnormal{l}_{1}$-norm prefers vectors with components of smallest non-zero magnitude, the $\mathnormal{l}_{0}$-norm considers only the least number of non-zero components. Equation \eqref{eq:crude_min_proof} can be further transformed into a linear programming problem:
\begin{equation}
\label{eq:min_proof}
\min_{\substack{\mathbf{y}\in \mathbb{R}^{m},\\ \mathbf{\mu}\in \mathbb{R}^{q}}} \mathbf{1^{\top}y} + \mathbf{1^{\top}t} \text{ subject to }
\begin{cases}
    \mathbf{y^{\top}G} = \mathbf{b}^{\top} + \mathbf{\mu^{\top}Q}\\
    -\mathbf{t} \leq \mathbf{\mu} \leq \mathbf{t}\\ 
    \mathbf{y}, \mathbf{t} \geq 0
\end{cases}.
\end{equation}


\subsection{Generating Shortest Counterexamples}


A similar analysis can be carried out to find the shortest counterexample. Motivated by the fact that some solutions to the dual problem $[\mathbf{y}^{\top}, \mathbf{\mu}^{\top}]$ are too complicated to comprehend the underlying information, we apply the same techniques as above to generate the shortest counterexample. We arrive at the linear program
\begin{equation}
\min_{\substack{\mathbf{y}\in \mathbb{R}^{m},\\ \mathbf{t}\in \mathbb{R}^{q}}} \mathbf{1}^{\top}\mathbf{y} + \mathbf{1}^{\top}\mathbf{t} \text{ subject to }
\begin{cases}
\mathbf{b}^{\top}  + \mathbf{\lambda}^{*\top}\mathbf{W}= \mathbf{y}^{\top}\mathbf{G} - \mathbf{\mu}^{\top}\mathbf{Q}\\
-\mathbf{t} \leq \mathbf{\mu} \leq \mathbf{t}\\
\mathbf{y}, \mathbf{t} \geq 0
\end{cases}
\end{equation}
The optimal solution $[\mathbf{y}^{*\top}, \mathbf{\mu}^{*\top}]$ of this problem can help determine additional equalities to potentially disprove the inequality.


\section{Examples}

Lastly, we will present three illustrative examples along with the corresponding outputs to demonstrate the capabilities of our prover.


\subsection{Proof of a von-Neumann-Type Inequality}

First, we show that a $5$-party Shannon-type inequality from Kaced's work \cite{Kaced_2013} is also a $5$-party von Neumann-type inequality. Kaced used this inequality to prove a $4$-party non-Shannon type inequality known as the Zhang-Yeung inequality, based on the classical copy-lemma method \cite{Zhang_1998} (see also \cite{Dougherty_2006}). It is given as
\begin{equation}
\label{eq:generalized_zy_inequality}
\begin{aligned}
    I(C;D\vert A)& + I(C;D\vert B) + I(C;D\vert E) + I(A;B)\\
    &+ I(C; E\vert D) + I(D; E\vert C)+ 3I(A,B; E\vert C,D)\\
    \geq I(C;D).
\end{aligned}
\end{equation}
To prove that Equation \eqref{eq:generalized_zy_inequality} is von-Neumann-type, the prover shows that the inequality is equivalent to
\begin{equation}
\label{eq:proof_vn_inq}
    \begin{aligned}
        &I(A;B\vert E) + I(A;E\vert C) + I(A;E\vert D) + I(A;E\vert B, C, D)\\
        &+I(B;E\vert C) + I(B;E\vert D) + I(B;E\vert A,C,D) \\
        &+ I(C;D\vert A,E) +I(C;D\vert B,E)+I(C;E\vert A,B,D)\\
        &+ I(D;E\vert A, B)\geq 0.
    \end{aligned}
\end{equation}
This can be verified by expanding Equation \eqref{eq:generalized_zy_inequality} and Equation \eqref{eq:proof_vn_inq} into their marginal entropy representation. However, coming up with such a solution as Equation \eqref{eq:proof_vn_inq}, is not obvious. This shows how the prover can save time when analysing information inequalities.


\subsection{Counterexample for Quantum Conditional Entropy}

In quantum information theory, the quantum conditional entropy can also get negative. Here, we verify that $S(A\vert B) \geq 0$ is in general not true for a tripartite quantum state $\rho_{ABC}$ using our prover. This shows how the generation of counterexamples can work. The prover first shows that
\begin{equation*}
S(A\vert B) \longrightarrow \textnormal{ Not provable by qITIP}
\end{equation*}
and suggests the following quantities altogether can be used to disprove it:
\begin{enumerate}
\item \label{1} $S(A,C) = S(A) + S(C)$,
\item \label{2} $S(A, B) + S(B, C) = S(B) + S(A,B,C)$,
\item \label{3} $S(A,B) + S(A,C) = S(A) + S(A,B,C)$,
\item \label{4} $S(A) + S(A,B,C) = S(B,C)$,
\item \label{5} $S(A,B) + S(B,C) = S(A) + S(C)$.
\end{enumerate}
In this case, we can construct a quantum state $\rho_{ABC}=\ket{\psi}\bra{\psi}_{ABC}$ that satisfies all the properties above as
\begin{equation*}
\ket{\psi}_{ABC} = \frac{1}{\sqrt{2}}(\ket{0, 1} + \ket{1, 0})_{AB}\otimes\ket{0}_{C}.
\end{equation*}
The corresponding entropic vector is
\begin{align*}
&\langle{S(A), S(B), S(C), S(A, B), S(A, C), S(B, C), S(A, B, C)}\rangle\\
&= \langle{\log{2}, \log{2}, 0, 0, \log{2}, \log{2}, 0}\rangle
\end{align*}
which violates the information inequality.


\subsection{Non-von-Neumann-type Inequalities}

The existence of unconstrained non-von-Neumann-type inequalities remains an open area of research. However, constrained non-von-Neumann-type inequalities are known \cite{Linden_2005, Cadney_2012, Majenz_2018} and in the following we analyse an example known as the Linden-Winter inequality. In \cite{Linden_2005}, it was proven that if a $4$-party quantum quantum state satisfies $I(A;C\vert B) = I(B;C\vert A) = I(A;B\vert D) = 0$, then
\begin{equation}
I(C;D) \geq I(A,B;C)
\end{equation}
holds and is a non-von-Neumann-type inequality. Our program then shows that, indeed, the constrained inequality is not von-Neumann-type. Showing by hand that the inequality is not von-Neumann-type was one of the crucial steps in the proof of \cite{Linden_2005}. Our prover provides a computer-assisted proof of this fact. Here, it is worth emphasizing again that a counterexample cannot be found for this example, despite the hints provided by our program.


\section{Conclusion}

Motivated by the classical ITIP, we presented a mathematical framework for proving and disproving information inequalities in quantum information theory based on linear programming techniques. We have implemented this framework in a Python package \cite{qitip_2024} and demonstrated its key functionalities. This can serve as a useful tool in the study of information inequalities. For the latest developments in this research area, see, e.g., \cite{Berta_2024, Bluhm_2025} and references therein. Moreover, as our prover can handle constrained inequalities as well, it also provides an important tool for further studying constrained non-von-Neumann-type inequalities (cf.\ \cite{Linden_2005, Cadney_2012, Majenz_2018}).


\section*{Acknowledgments}

We acknowledge funding by the European Research Council (ERC Grant Agreement No. 948139) and support from the Excellence Cluster - Matter and Light for Quantum Computing (ML4Q).

\clearpage
\bibliographystyle{IEEEtran}
\bibliography{main}

\end{document}